\documentclass[11pt]{article}
\usepackage{vietnam,epsfig}

\newlength{\ogltemp}

\def\met{\settowidth{\ogltemp}{\big\slash}%
\hspace{0.5mm}\big\slash\hspace{-1.4\ogltemp}{E_T}%
}
\def\D0{${\rm D}{\rm O\!\!\!\!\hspace{0.5pt}\protect\raisebox{0.2ex}{/}}$}

\def\be{\begin{equation}}
\def\ee{\end{equation}}
\def\bea{\begin{eqnarray}}
\def\eea{\end{eqnarray}}

\def\gev{\,\,\mathrm{GeV}\,}
\def\tev{\,\,\mathrm{TeV}\,}

\def\gevc{\,\,\mathrm{GeV}/c\,}
\def\gevm{\,\,\mathrm{GeV}/c^2\,}
\begin{document}
\vspace*{4cm}
\title{Searching for SUSY at the Tevatron}

\author{ D. Bortoletto }

\address{Physics Department, Purdue University, 525 Northwestern Avenue,\\
West Lafayette, IN, 47906, U.S.A.\\
{\rm Representing the CDF and \D0 collaborations}}

\maketitle\abstracts{ An overview of recent experimental searches
for SUSY particles is presented. These searches are based on data
collected by the CDF and the \D0 experiments operating at the
Fermilab Tevatron proton-antiproton collider with $\sqrt{s}$ =
$1.96$ TeV. The review focuses on searches for squarks and gluinos
in final states with missing transverse energy and jets. Emphasis
will be given to the search for the gluino decaying into a sbottom
and $b$ quark with each sbottom decaying into a $b$ quark and a
neutralino. This scenario yields events containing 4 $b$-jets and
missing transverse energy.}

\section{Introduction}

Despite of its extraordinary success, the Standard Model (SM) is
believed to be an effective "low energy theory" of a more
fundamental theory. One attractive extension of the SM is
Supersymmetry (SUSY)~\protect\cite{Martin:1997ns}, a spacetime
symmetry that relates bosons to fermions and introduces for each
SM particle a SUSY partner. SM and SUSY particles carry a value of
+1 and -1 for R$-$parity respectively. Therefore in R$-$parity
conserving theories SUSY particles are produced in pairs and they
decay into lighter SUSY particles until they decay to SM particles
and the lightest supersymmertic particle (LSP) which is stable and
it would escape detection. One of the most interesting LSP
candidate is the lightest neutralino $\tilde{\chi}^0_1$ which
might also explain dark matter in the universe. In this paper, we
will consider only R-parity conserving searches where the presence
of missing transverse energy is a powerful signature of SUSY.

At hadron collider we expect that gluinos and squarks would be
copiously produced if they are sufficiently light. If squarks are
lighter than gluinos  then the squarks can decay to
$\tilde{q}\rightarrow q\tilde{\chi}^0_1$. In most SUSY models, it
is expected that the partners of the four lightest quark will have
similar masses. On the contrary, the partners of the heaviest
quarks could be the lightest squarks. The states $\tilde{q}_L$ and
$\tilde{q}_R$ are the partners of the left-handed and right-handed
quarks, which mix to form mass eigenstates $\tilde{q}_{1,2}$. The
stop is expected to be light because of the large mass of the top
quark. For scenarios with large $\tan\beta$ (the ratio of the
vacuum expectation values of the two Higgs fields), the mixing can
be quite substantial in the sbottom
sector~\protect\cite{Bartl:1994bu}, so that the lighter sbottom
mass eigenstate (denoted by $\tilde{b}_1$), could also be
significantly lighter than other squarks.

In this paper we review current results from a gluino-squark
search from the \D0 collaboration and a gluino-sbottom search from
the CDF collaboration. The trigger for both searches required jets
and significant missing energy.
%However there are many sources of $\met$ in an event such as neutrinos and detector which yield to
%large Standard models background that must be well understood
%before setting limits.

\section{Gluino and Squark searches}

The \D0 collaboration has performed a search for gluinos and
squarks using 85 pb$^{-1}$ collected in run II of the Tevatron.
The selection cuts developed by the \D0 collaboration aimed at
enhancing the SUSY signal and rejecting the large SM backgrounds.
The analysis requires two jets in the event. The leading jet is
required to be central $(|\eta|<0.8)$ and to have transverse
momentum above $60\gevc$. Events with an isolated electron or a
muon with momentum above 10 GeV/c were rejected to decrease the
electro-weak background. The QCD multi-jet background was
significantly reduced by rejecting events were the missing
transverse energy is parallel or antiparallel to a jet by
requiring the minimum $\Delta\phi(\met,jet) > 30^{\circ}$ and
maximum $\Delta\phi(\met,jet) < 165^{\circ}$. The final selection
criteria were optimized by minimizing the expected cross section
upper limit and requiring $\met>175 \gev$ and the transverse
energy of the jets $H_T > 275 \gev$. The missing transverse energy
($\met$) is defined as the negative vector sum of the transverse
energy in the electromagnetic and hadronic calorimeter.

In order to extract the SUSY signal, the SM background must be
calculated.  The Electro-weak decays of W and Z boson were
generated with the ALPGEN program, interfaced with PYTHIA for the
simulation of initial and final state radiation and jet
hadronization. The most significant backgrounds is due to
$Z\rightarrow \nu \bar{\nu}$ + 2 jets. The QCD background was
evaluated from the data using events with low $\met$ and for
$\met>175 \gev$ was found to be negligible.

\begin{figure}[d01]
\begin{center}
\includegraphics*[width=7cm]{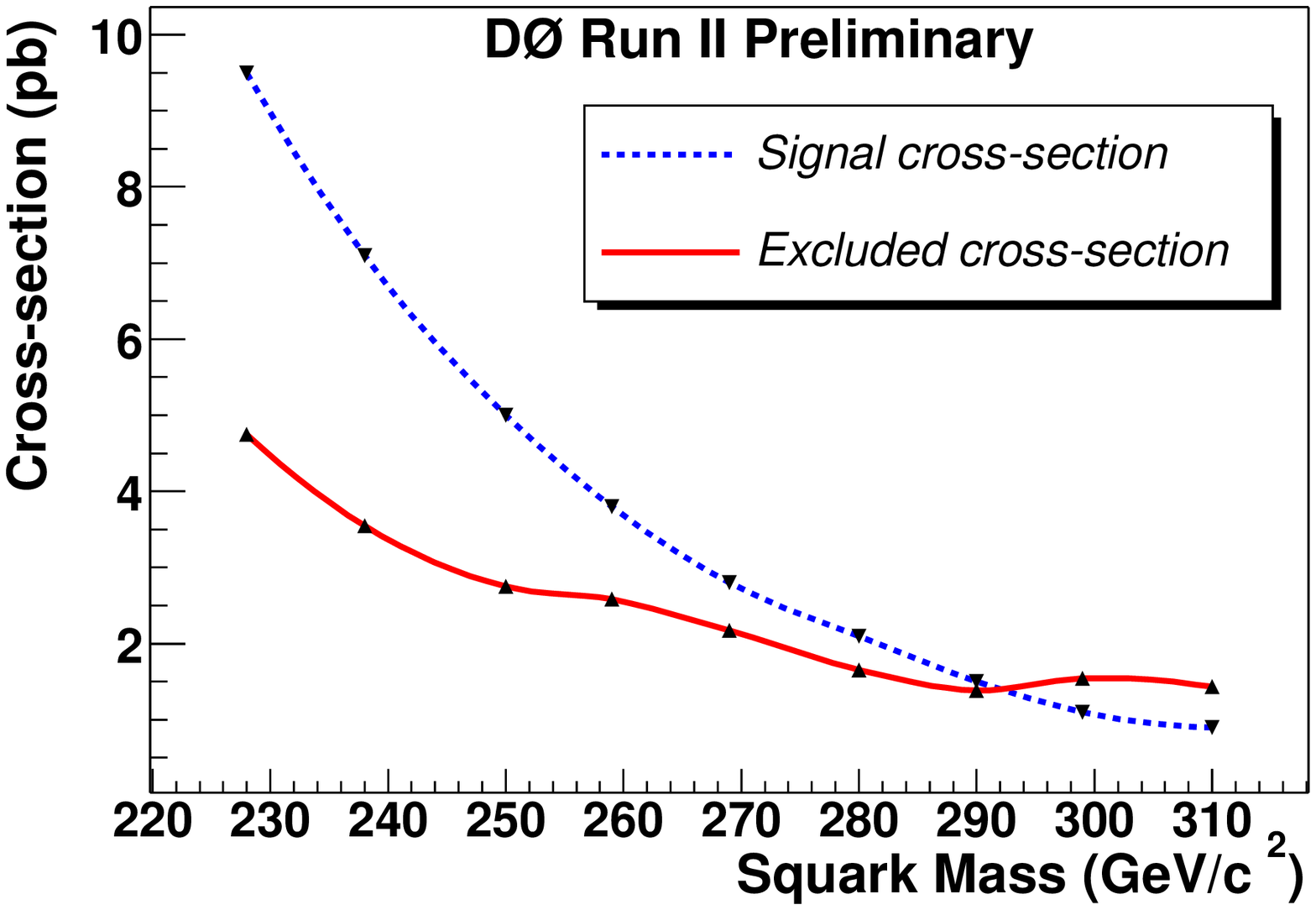}

\includegraphics*[width=7cm]{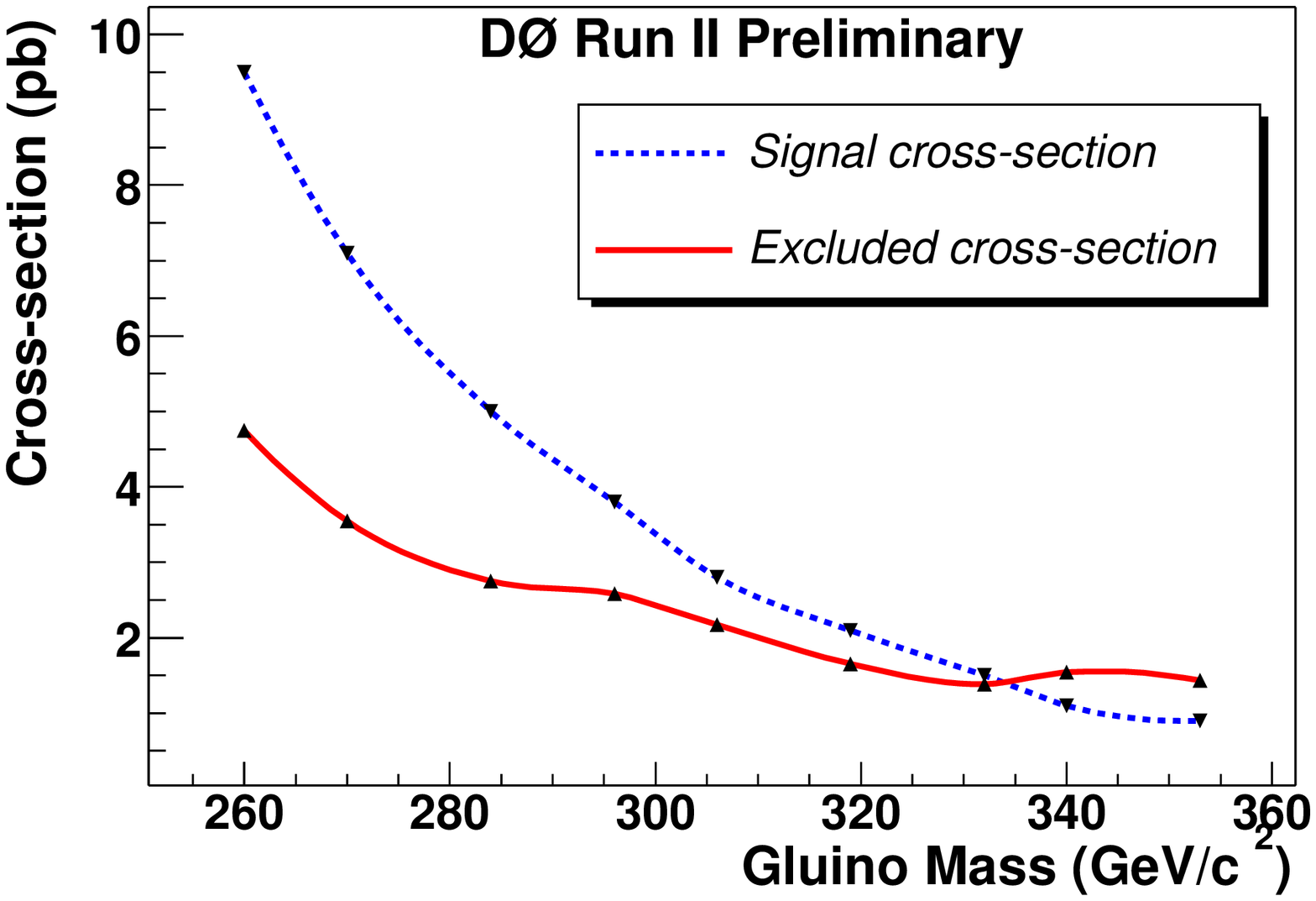}

\caption{squark and gluino production cross section upper limit as
a function of the average squark (top) and gluino (bottom) masses
and theoretical expectations for $m_0=25 GeV/c^2$, $A_0=0$ , $\tan
\beta=3$ and $\mu<0$} \label{d01}
\end{center}
\end{figure}

The simulation of the production of squarks and gluinos was done
with the PYTHIA Monte Carlo in the framework of mSUGRA. The
K-factor for the various production processes was determined by
Prospino\cite{hep-ph9611232}. The input parameters were chosen at
the boundary of the Run I exclusion domain: $m_{0}=25 \gevm$,
$\tan{\beta}$=3, $A_{0}=0$, $\mu<0$, and $m_{1/2}$ in the range
100 to 140 $\gevm$ in 5 $\gevm$ steps. Only squarks belonging to
the first two generation were considered. The signal efficiency
after these selection varies between 2 and 7 \% as the $m_{1/2}$
increases from 100 to 140 $\gevm$. The main systematic
uncertainties are due to the 6.5 \% uncertainty in the integrated
luminosity and the uncertainty jet energy scale in the data and in
the Monte Carlo which yields a $^{+20}_{-15}$\% relative
uncertainty on the signal efficiency and a $^{+77}_{-43}$\%
uncertainty on the SM background prediction. The systematic error
on the signal and the SM background cross sections are about 10 \%
and 8\% respectively. \D0 observes four events while $2.7 \pm
0.95$ events were expected from SM backgrounds. Taking into
account the signal efficiency, the systematic uncertainties and
the integrated luminosity of $85.1 \pm 5.5$ pb$^{-1}$, cross
section limits were set as a function of $m_{1/2}$ for the mSUGRA
model line defined above. The results are displayed in
Fig.~\ref{d01}. This limits exclude a gluino mass of 333 $\gevm$
for a squark mass of 292 $\gevm$. The corresponding most probable
gluino-mass limit expected in this analysis is 338 $\gevm$.  The
new analysis has already pushed the 95 \% C.L. slightly beyond the
Run I limits \cite{run1ana}.

\section{Gluino-Sbottom searches}

The CDF collaboration has conducted a search for gluino-sbottom
production using 156 pb$^{-1}$ of data collected during Run II of
the Tevatron. The analysis assumes a scenario where the sbottom is
lighter than the gluino and relies on the large gluino pair
production cross-section. Next to leading order calculation (NLO)
using PROSPINO~\protect\cite{hep-ph9611232} predicts a
cross-section of $2.04$~pb for a gluino of mass $240\gevm$ at
$\sqrt{s}=1.96\tev$, which is large compared cross-section of
$0.072$~pb for sbottoms of the same mass.  R-parity conservation
is assumed and a neutralino with a mass of $60\gevm$ is supposed
to be LSP. A SUSY scenario where gluinos are pair-produced and
then decay $100\%$ into sbottom bottom
($\tilde{g}\rightarrow\tilde{b}_1b$), followed by the sequential
decay of the sbottom in bottom and lightest neutralino
($\tilde{b}_1\rightarrow b\tilde{\chi}_1^0$) is assumed. Since the
neutralinos escape detection, this leaves a signature of four
b-jets and missing transverse energy ($\met$).

The data sample is collected with a trigger that requires $\met
\equiv |\vec{\met}| \ge 35$~GeV and two jet clusters. Even after
stringent selections the data is dominated by QCD multijet, with
the large $\met$ resulting from jet mismeasurements or from
semileptonic b-decays in which the neutrino escapes detection. In
both cases the $\met$ is aligned with mismeasured jets or b-jets.
The $\met$ in signal originates form the neutralinos from the
sequential decay of the gluinos and is therefore not correlated to
the jets.  By requiring that the minimum
$\Delta\phi{(\met,jet)}>40^{\circ}$, the SM background can be
effectively reduced while keeping a large signal acceptance. A
secondary vertex tagging algorithm is applied to identify b-jets
and to reduce the background further.

Understanding of the SM background is achieved by studying control
regions neighboring the signal box.  The regions are defined in
terms of $\met$ and the presence of high momentum isolated
leptons. We expect our signal to have large $\met$ and no leptons.
The regions with low $\met$ ($35\gev < \met < 50\gev$) with and
without a lepton serve as control regions for the QCD multijet
background. The control region with $\met>50 \gev$ and containing
high transverse momentum isolated leptons is used to check the top
and W/Z+jets/Diboson background.

ALPGEN~\protect\cite{alpgen} in combination with the
HERWIG~\protect\cite{herwig} event generator was used to estimate
the acceptance of the W and Z boson background. The cross sections
at NLO were obtained using the MCFM~\protect\cite{mcfm} program.
The top contributions and the QCD heavy flavor background were
predicted using the PYTHIA~\protect\cite{pythia} event generator.
The QCD light quark contribution which originates from light quark
jets being mis-identified as heavy flavor jets was estimated using
a parametrization of the fake tag rate obtained from data. Various
distributions in the control regions have been studied and found
to be in agreement with observations.

The signal predictions were computed using the
ISAJET~\protect\cite{isajet} event
generator with the CTEQ5L %~\protect\cite{CTEQ5L}
parton distribution functions.

\begin{figure}[t]
\begin{center}
\includegraphics*[width=7 cm]{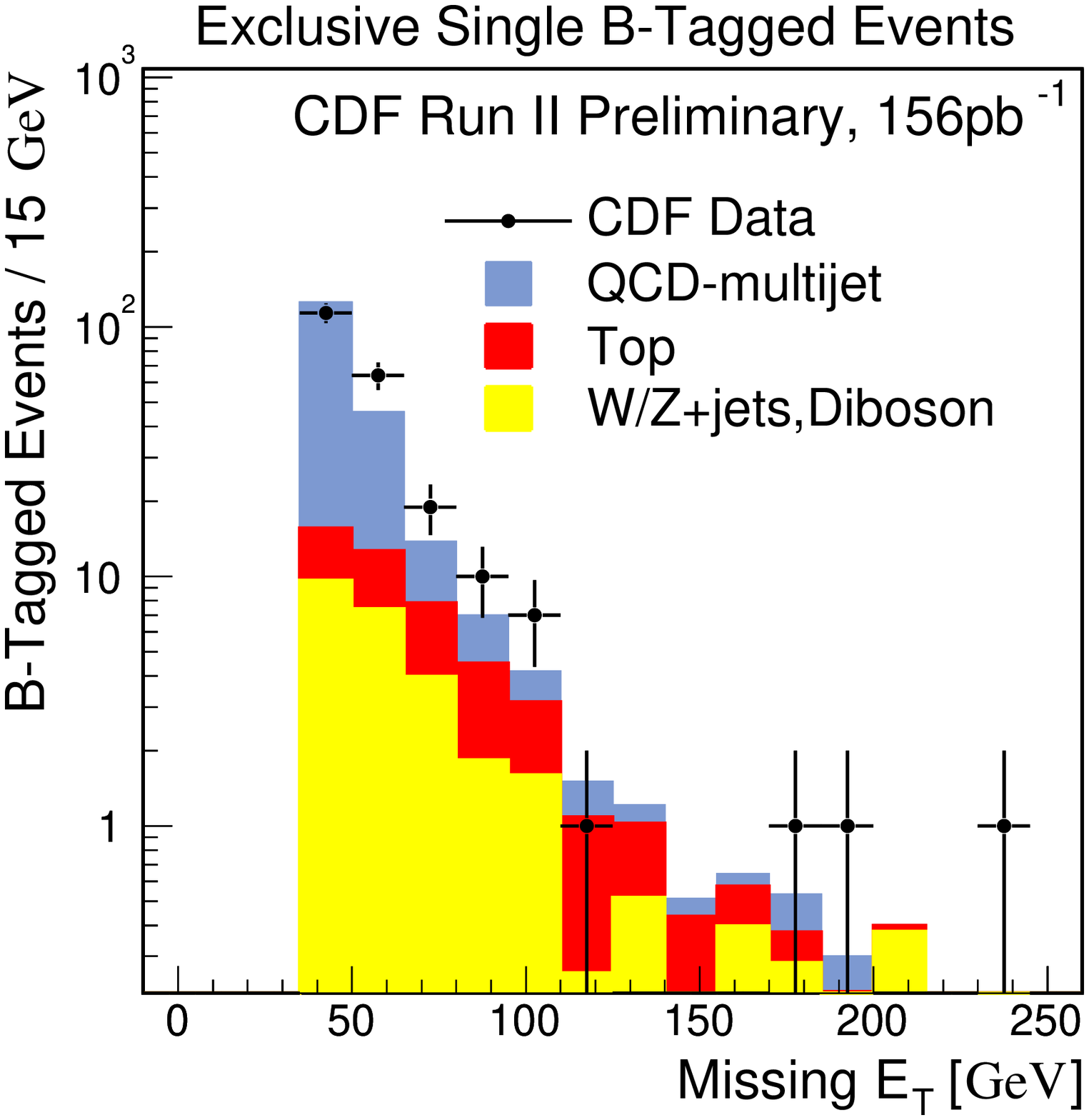}
\includegraphics*[width=7 cm]{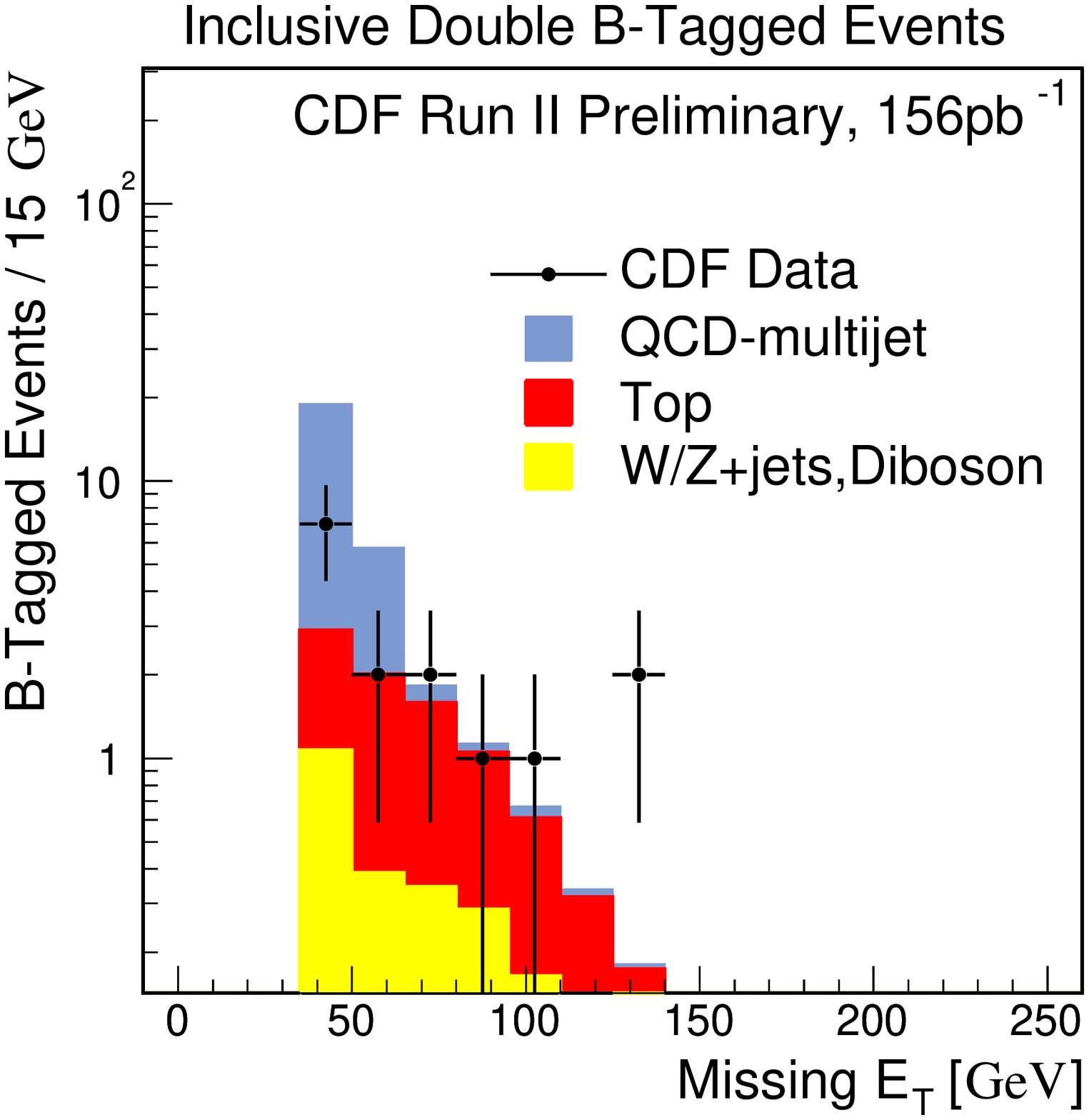}
\caption{$\met$ spectrum after vetoing events with high $P_t$
isolated leptons, for exclusive single tagged events (left) and
inclusive double tagged events (right).} \label{fig:spectrums}
\end{center}
%\label{fig:spectrums}
\end{figure}

CDF performs two analysis, one using exclusive single b-tagged
events and the other requiring inclusive double b-tagged events.
The single b-tag analysis is expected to have a better reach for a
nearly mass degenerated gluino-sbottom scenarios since b-jets from
gluino decays are very soft and less likely to be tagged. The
double tag analysis suppresses the background more effectively and
it is expected to perform better in the other kinematic regions.
Fig.~\ref{fig:spectrums} shows the $\met$ spectrum for both cases.
The best signal sensitivity was achieved by requiring
$\met>80\gev$, which was optimized using signal MC.

The signal acceptance systematic uncertainty for the exclusive
single tag analysis (16.5\% in total) was dominated by jet energy
scale (10\%), modeling of initial and final state radiation
(7.5\%), b-tagging efficiency (7\%), luminosity (6\%), Monte Carlo
statistics (3\%), trigger efficiency (2.5\%), parton distribution
functions (2\%), and lepton veto (2\%). The uncertainties for the
inclusive double tag analysis were very similar, except the
b-tagging efficiency systematics was increased.

The signal region was only analyzed after all the background
predictions and selection cuts were finalized. Twenty one
exclusive single b-tagged events were observed, which is in
agreement with SM background expectations of $16.4 \pm 3.7$
events. Requiring inclusive double b-tag we observed four events,
where $2.6\pm0.7$ were expected, as summarized in
Table:\ref{tab:table1}.

\begin{table}[h]
\begin{center}
\caption{
Number of expected and observed events in the signal region. % with a $\met>80$~GeV.
}
\begin{tabular}{|l|c|c|}
\hline
Process                         &Exclusive Single B-Tag & Inclusive Double B-Tag \\
\hline\hline
W/Z+jets/Diboson & $    5.66 \pm 0.76\rm{(stat)}  \pm  1.72(sys)     $ & $  0.61 \pm 0.21\rm(stat) \pm 0.19(sys)         $ \\
TOP              & $    6.18 \pm 0.12\rm{(stat)}  \pm  1.42(sys)     $ & $  1.84 \pm 0.06\rm(stat) \pm 0.46(sys)         $ \\
QCD multijet     & $    4.57 \pm 1.64\rm{(stat)}  \pm  0.57(sys)
$ & $  0.18 \pm 0.08\rm(stat) \pm 0.05(sys)         $ \\ \hline
\hline Total predicted  & $    16.41 \pm 1.81\rm{(stat)} \pm
3.15(sys)     $ & $  2.63 \pm 0.23\rm(stat) \pm 0.66(sys)
$ \\  \hline Observed         & $    21                       $ &
$  4 $ \\ \hline
\end{tabular}
\label{tab:table1}
\end{center}
\end{table}

\begin{figure}[t]
\begin{center}
\includegraphics*[width=8 cm]{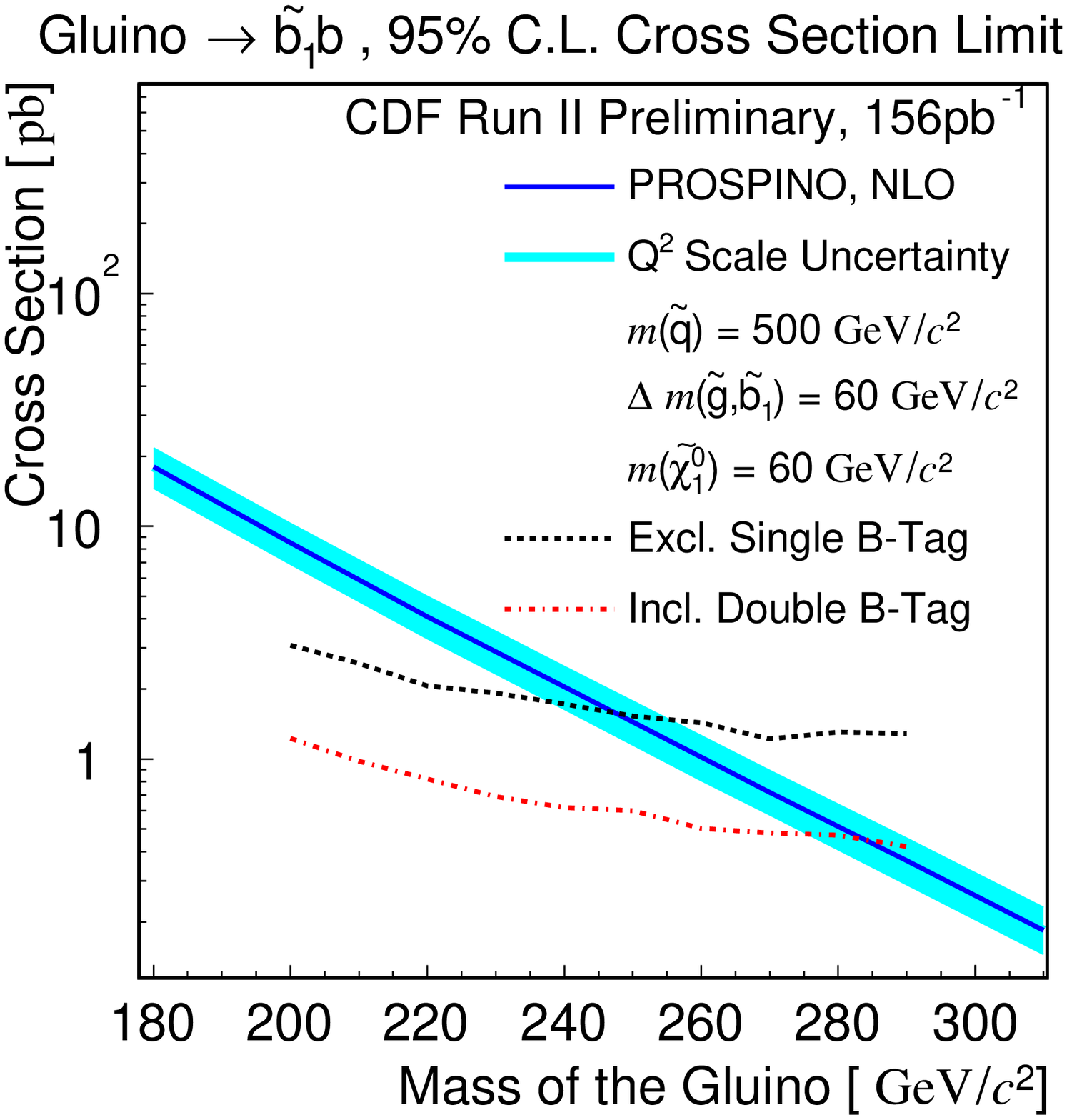}
\includegraphics*[width=8 cm]{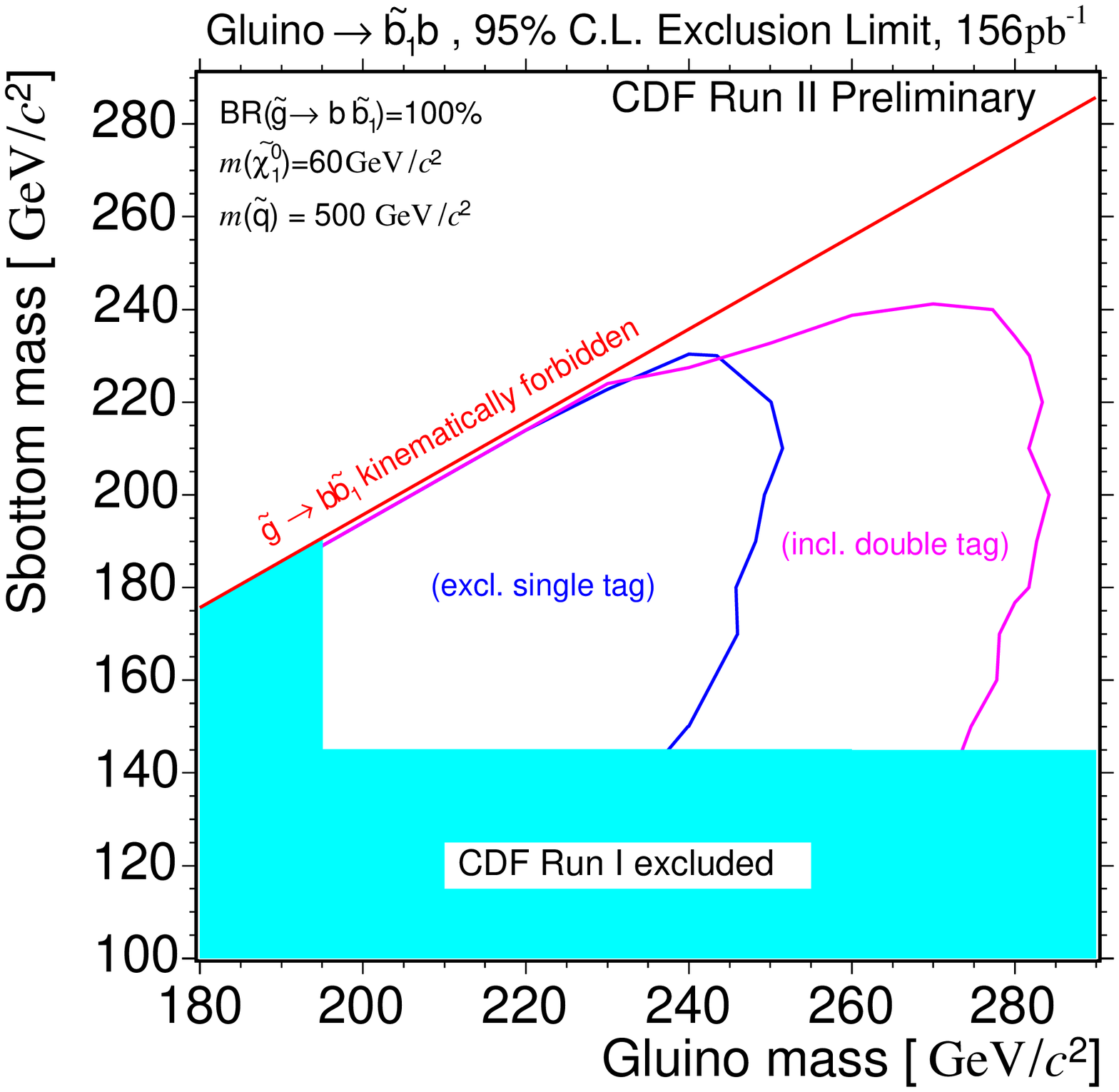}
\caption {
           Top: 95\% C.L. upper cross section limit as a function of the gluino mass and
           for a constant mass difference $\Delta m(\tilde{g},\tilde{b}_1) =60\gevm$ between the gluino and sbottom.
           Bottom: 95\% C.L. exclusion contours in the $m(\tilde{g})$ and $m(\tilde{b_1})$ plane
           obtained requiring exclusive single b-tagged events and inclusive double b-tagged events.
} \label{fig:exclusion}
\end{center}
\end{figure}

Since no evidence for gluino pair production with sequential decay
into sbottom-bottom was found, an upper limit cross sections at
$95\%$ C.L. was computed using the Bayesian likelihood method.
cross section limit and the exclusion plot as a function of the
gluino and squark masses are shown Fig.~\ref{fig:exclusion}). The
results improve the existing limits significantly.

% \begin{figure}[t]
% \begin{center}
% \caption
% {95\% C.L. exclusion contours in the $m(\tilde{g})$ and $m(\tilde{b_1})$ plane.
% }
% \includegraphics*[width=5cm]{cfig8.eps}
% \end{center}
% \label{fig:exclusion}
% \end{figure}

\section{Conclusion}

Searches for new physics have been performed by CDF and \D0. Even
with a fraction of the data that it will be collected by the end
of Run II of the Tevatron the two experiments are significantly
improving current limits.  No evidence for squark-gluino or for
gluino-sbottom production has been found.

\section{Acknowledgments}

I would like to thank the organizing committee for this exciting
and enjoyable meeting and the great introduction to Vietnam.

%We thank ....

\bibliographystyle{plain}

\end{document}